\documentclass{article}
\usepackage{amsmath,amsfonts,amssymb}
\usepackage[utf8]{inputenc}
\usepackage{authblk}
\usepackage{natbib}
\usepackage{xcolor}
\usepackage[version=4]{mhchem}
\usepackage{siunitx}
\usepackage{placeins}
\usepackage{fancyhdr}
\DeclareSIUnit\Molar{M}
\usepackage{caption}
\usepackage[normalem]{ulem}
\usepackage{lineno}
\usepackage{url}
\usepackage{bm}
\usepackage{bbm}
\usepackage{float}
\usepackage{mathtools}
\usepackage[toc,page]{appendix}
\usepackage{needspace}
\usepackage{xpatch}
\usepackage{gensymb}
\usepackage{cancel}
\usepackage{stackrel}
\usepackage{ifthen}
\usepackage{chngcntr}
\usepackage{ntheorem}
\usepackage{graphicx}
\usepackage{cancel}
\usepackage[colorlinks=true, allcolors=blue]{hyperref}

\newcommand{\pare}[1]{\left(\, #1 \, \right)}

\newcommand{\vF}{\vec{F}}

\newcommand{\Cell}[2]{{#1}_{#2}} 

\newcommand{\Co}[4]{\Gamma^{\Cell{#1}{#2}}_{\Cell{#3}{#4}}}

\newcommand{\W}[4]{W^{\Cell{#1}{#2}}_{\Cell{#3}{#4}}}

\newcommand{\cK}{{\mathcal K}}

\newcommand{\cI}{{\mathcal I}}

\newcommand{\vV}{\vec{V}}

\newcommand{\vpsi}{\vec{\psi}}

\title{Distinct inhibitory connectivity motifs trigger distinct forms of anticipation in the retinal network}

\date{}

\author[1,2,3]{Simone Ebert}
\author[1]{Bruno Cessac}
 
\affil[1]{Université Côte d’Azur, Inria Biovision Team and Neuromod Institute}
\affil[2]{Sorbonne Université, INSERM, CNRS, Institut de la Vision, 17 rue Moreau, F-75012 Paris, France}
\affil[3]{Hertie Institute for AI in Brain Health, University of Tuebingen, Germany}
\affil[*]{simone.ebert@uni-tuebingen.de}

\begin{document}
\maketitle

\begin{abstract}

Motion is an important feature of visual scenes and retinal neuronal circuits selectively signal different motion features. It has been shown that the retina can extrapolate the position of a moving object, thereby compensating sensory transmission delays and enabling  signal processing in real-time. Amacrine cells, the inhibitory interneurons of the retina, play essential roles in such computations although their precise function remain unclear. Here, we computationally explore the effect of two different inhibitory connectivity motifs on the retina's response to moving objects: feed-forward and recurrent feed-back inhibition. We show that both can account for motion anticipation with two different mechanisms. Feed-forward inhibition truncates motion responses and shifts peak responses forward via subtractive inhibition, whereas recurrent feedback coupling evokes, via divisive inhibition, excitatory and inhibitory waves with different phases that add up and shift the response peak. A key difference between the two mechanisms is how the peak response scales with the speed of a moving object. Motion prediction with feedforward circuits monotonically decreases with increasing speeds, while recurrent feedback coupling induces tuning  curves that exhibit a preferred speed for which motion prediction is maximal.

\end{abstract}

\section{Introduction}
\label{Introduction}


A visual scene is constantly in motion, not only because our external environment contains moving objects, but also because ourselves and our eyes are constantly moving. Already in the retina, several cell-types are specialized to detect a variety of  motion features  such as the global motion of a visual scene versus the local motion of an object \cite{Olveczky2003, Kim2017}, looming objects \cite{Munch2009, Kim2020, Jo2023} or the direction of motion \cite{Wei2018, Demb2007}.

In addition to detecting these features, the brain has to respond in real-time. To compensate for delays in neuronal responses, retina has been shown to predict the trajectory of a moving object \cite{Berry1999}: When a bar moves across the receptive field of retinal ganglion cells (RGCs), the peak-firing rate of the cell occurs earlier as if the bar is flashed above the receptive field center. This peak shift of the response has been coined "motion anticipation" and implies that already in the inner retina, cells form a prediction of the future position of a moving object \cite{Palmer2015}.

This predictive capacity is generally believed to be implemented via suppression of negative feedback, such as a reduction in response gain (gain control, \cite{Berry1999}), which can also account for more complex motion extrapolation, such as tracking the position of a moving object in a 2D-plane \citep{Leonardo2013}. However, the principle of gain control is a rather broad and phenomenological description. It could have many underlying biophysical mechanisms. Several other studies provided more mechanistic explanations and showed how amacrine inhibition can contribute to motion anticipation through spatially and temporally displaced feedforward inhibition \cite{Johnston2015}, or by forming an anticipatory wave ahead of the stimulus via gap junctions  \citep{Liu2021, Trenholm2013a, Trenholm2013b, Kuo2016} or lateral recurrent feedback \citep{Souihel2021}. Further mechanisms for anticipation have been proposed to act additionally in the visual  cortex \cite{Benvenuti2020, Emonet2025}. 
\\

In order to anticipate a moving object, at the right position at the right time, its speed needs to be estimated.
A wide range of retinal cell types have been shown to be tuned to the velocity of a moving object \citep{Jacoby2017, Ravello2019, Strauss2022, Mani2023, Matsumoto2025}, and that RGCs encode object speed for motion processing \citep{Deny2017}. A recent study has shown that the amplitude response of some cells signaling motion direction is velocity invariant, while other cells exhibit a tuning to velocity. These two response behaviors are implemented by distinct inhibitory circuits \cite{Summers2022}. In addition, different cell types have different scaling properties relating anticipation and the speed of a moving bar. Some cells exhibit a tuning to a "preferred" speed at which anticipation is maximal, while others maintain a stable anticipation in a wide speed range, e.g. $0.1-1.0$ mm/s in \cite{Berry1999} or \cite{Johnston2015}. 
\\

Here, we computationnally study  how two different connectivity motifs of inhibitory interneurons impact the network response to static and dynamic input. In particular, we study the effect of (1), \textit{feed-forward inhibition} and (2), \textit{recurrent feed-back inhibition}. 
First, we show how these two different connectivity motifs affect basic response properties such as the steady-state response to a step stimulus and the shape of the response to a flash.
We then show how the two connectivity motifs can implement motion anticipation in response to a moving bar and observe that they act at two different stages in the retinal network: Feedforward inhibition induces a peak shift at the level of RGCs while recurrent feedback inhibition acts already at the level of bipolar and amacrine cells. 
Finally, we investigate how anticipation scales with the speed of the moving bar. We show that feedforward connectivity acts strongest on slower speeds, while feedback connectivity gives rise to a tuning curve with a preferred speed for which anticipation is maximal. This tuning depends on the strength of recurrent connectivity, which could allow the network to extend the range of motion anticipation and maintain a stable anticipation time via dynamic adaptation. 
\\


\section{Results}
\label{Results}

\subsection{A retinal model for Motion Anticipation via Inhibition}\label{Sec:Model}

To study how feedforward and recurrent feedback amacrine connectivity impacts retinal responses to static and dynamic stimuli, we designed a 1-D network model consisting of 3 layers to simulate retinal processing of an incoming stimulus (Figure \ref{fig:1}). Individual bipolar cells (BCs), amacrine cells (ACs) and retinal ganglion cells (RGCs) are simulated as point neurons and are characterized by their voltage response, $V(t)$. 

The equations describing the model are presented in the Supplementary section \ref{Sec:Equations}. To simulate retinal processing, the spatiotemporal stimulus $s(x,t)$ is first convolved with a spatiotemporal kernel, with a characteristic time of integration $\tau_{RF}$, to simulate how the Outer-Plexiform Layer (OPL) transforms a visual scene into a voltage response in bipolar cells. This mathematically describes the receptive field (RF) of the respective BC, in the context of linear response.
This response, $V_{drive}(t)$ (eq. \eqref{eq:V_drive}), is purely evoked by the stimulus.  The spatial profile is a Gaussian filter corresponding to the receptive field of the bipolar cell $i$ centered at its position $x_i$. Here, we simulate only the positive center input that comes from photoreceptor input, while surround inhibition  from horizontal cells in the OPL \cite{Diamond2011} is not taken into account for simplicity. All surround effects in the model are mediated by BC-AC interactions. 
The temporal kernel sums up transmission delays in this first retinal processing step. 

The second layer of the model simulates BCs and ACs interactions and is featured as a dynamical system, where the voltage response of each cell type has a characteristic time constant $\tau_B$ (for BCs) or $\tau_A$ (for ACs) (see eq. \eqref{eq:dynamical_systemc4}). The network consists of two sub-layers of $N$ regularly spaced BCs, with index $i = 1...N$, and  $N$ ACs,  $j = 1...N$, which share the same horizontal spatial position $x_{i} = x_j$. The distance between two neighbours cells is noted $\delta$ (expressed in mm). 

The connectivity is simulated through the connectivity matrices $\Gamma^{B}_{A}$ and $\Gamma^{A}_{B}$, which define the connections from BCs to ACs and from ACs to BCs respectively. Each BC is reciprocally connected to neighboring ACs such that the connectivity between the two layers obeys $\Gamma^{B}_{A} = \Gamma^{A}_{B}$ so that $\Gamma^{B}_{A},\Gamma^{A}_{B}$ have the same eigenvalues $\kappa_n$, and eigenvectors $\vpsi_n$, $n=1 \dots N$.  Each BC $i$ projects onto ACs $j = i-1$ and $j = i+1$ and vice versa so that $\Gamma^{B_{i}}_{A_{i-1}} = \Gamma^{B_{i}}_{A_{i+1}} = 1$ and $0$ otherwise.  For these "nearest-neighbours" connectivity matrices, each eigenvector is characterized by a space scale $\frac{N\delta}{n}$ \citep{Souihel2021}. The connections from BCs to ACs are excitatory and have synaptic weight $w^{+} \geq 0$ while connections from ACs to BCs are inhibitory and have synaptic weight $-w^{-} \leq 0$.  These inhibitory connections back from AC to BC are referred to as recurrent connectivity (Figure \ref{fig:1} \textbf{A}, dark lines). This connectivity scheme is clearly simplified compared to biological lateral connectivity in the retina but affords mathematical analysis.

In the third layer of the model, retinal ganglion cells (RGCs) are pooled over the voltage responses of bipolar and amacrine cells within their receptive field through weighted linear synapses. The weight $w^{B_i}_{G_k} \geq 0 $ describes excitatory connections from BC $i$ to RGC $k$ and the weight $w^{A_i}_{G_k} \leq 0 $ describes inhibitory connections from AC $i$ to RGC $k$. Both weights are scaled by a Gaussian distribution depending on the distance of the cell $i$ from the RF-center of the RGC $k$ (see Equations \eqref{eq:wxb}, \eqref{eq:wxa} in the Supplementary Material). These inhibitory connections are referred to as feedforward connectivity (Figure \ref{fig:1} \textbf{A}, bright lines).
 We note $\W{B}{}{G}{}$ (resp. $\W{A}{}{G}{}$) the connectivity matrix from BCs to RGCs (ACs to RGCs).

The dynamics of the RGCs layer is described by $N$ differential equations. They integrate their input with a shared time constant $\tau_{G}$ (eq. \eqref{eq:Vg}). Their voltage response is passed through a non-linear function $N_{G}$ to simulate a firing rate $R_{G}(t)$.

The dynamics of the model can be summarized and tuned by the time constants $\tau_{RF},\tau_B, \tau_A, \tau_G$, and the synaptic weights $w^+,w^-,w^B_G, w^A_G$ of the model. As shown in \cite{Souihel2021,Kartsaki2024}, some of these parameters are linked to the recurrent connections between BCs and ACs, and can be combined to define dimensionless parameters that have a deep impact on global dynamics. Especially, the product:
\begin{equation}\label{eq:eta}
 \eta=w^-\,w^+\,\tau_B\,\tau_B,   
\end{equation}
quantifies the feedback effect between BCs and ACs, whereas the parameter:
\begin{equation}\label{eq:mu}
   \mu=w^-\,w^+\tau^2;\quad \frac{1}{\tau} \,=\, \frac{1}{\tau_A} \,-\, \frac{1}{\tau_B},
\end{equation}
determines the appearance of oscillations and waves in dynamics \cite{Souihel2021,Kartsaki2024}.

A frequently observed property of the circuits in the retina is that synaptic connections dynamically adapt during stimulation \cite{vonGersdroff1997,Oesch2011, Singer2006,Jarsky2011}. Synaptic weights are not static but tuned to the visual input. To account for this effect in this computational study we will modulate the parameters $\eta,\mu$ through the strength of inhibitory connectivity weight $w^{-}$.

We will explore the effects of inhibitory feedforward  and inhibitory feedback connections separately. For simulations with feedforward connectivity, we set $w^{-} = 0 $ to remove feedback inhibition. For simulations with feedback connectivity, we set $w^{A}_{G} = 0$. 


\subsection{General response properties of the different connectivity motifs}

First, we study how feedforward and recurrent feedback connectivity lead to different network responses to simple static stimuli, such as a sustained light step and a short flash (impulse).
\\

\begin{figure}[h!]
\includegraphics[width=\textwidth,keepaspectratio]
{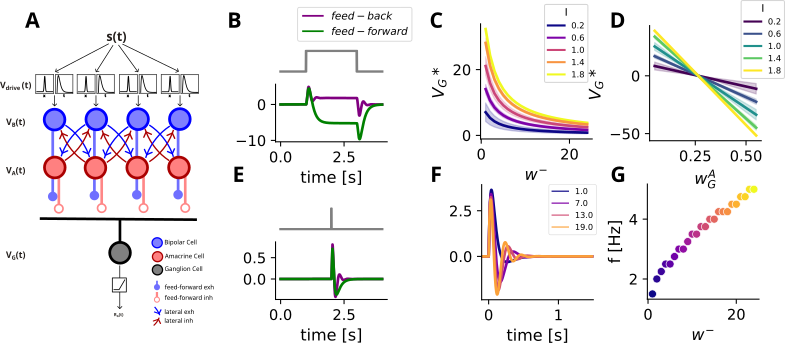}
\caption{\textbf{Schematic description of the model and its general response properties.} 
\textbf{A.} The stimulus $s(x,t)$ is fed into a convolution layer that simulates the transformation of the visual input into a neuronal voltage response, $V_{drive}(t)$, for each BC in the network. This convoluted signal is then fed into a network of BCs and ACs, which are reciprocally connected and pass the synaptic signals $V_B(t)$ and $V_A(t)$ on to neighboring cells of the other type. A third layer of RGCs pools over BCs within their receptive field and integrate their response $R_B(t)$ into their voltage $V_G(t)$. This voltage is transformed into a firing rate response $R_G(t)$ after rectification. 
\textbf{B.} Example of step response with both connectivity motifs, feed-back (purple) and feed-forward (green) inhibition. They evoke a similar transient response at the onset of the stimulus, which then decays to a rest state, that differs between motifs.
\textbf{C.} Rest state potential for constant and spatially homogeneous inputs of different amplitudes across recurrent feed-back inhibitory strength $w^-$.
\textbf{D.} Rest state potential for constant inputs of different amplitudes across feedforward inhibitory strength $w^G_A$.
\textbf{E.} Example of impulse response with both connectivity motifs. 
\textbf{F.} Impulse responses for different recurrent inhibitory strengths $w^-$. 
\textbf{G.} Leading Frequency of impulse response varies with  recurrent inhibitory strength $w^-$. Same color legend as in F.}
\label{fig:1}
\end{figure}

In response to a static and spatially homogeneous full-field input, both inhibitory motifs lead to a transient response at the onset of the stimulus, which then decays to a rest state (Figure \ref{fig:1} B). This rest state depends on the input strength and its analytic form can be easily computed. It strongly differs between the two connectivity motifs. We note $\vF$ the constant stimulus vector, such that a BC with index $i$ receives an input $F_i$ (with unit mV/ms). 

%
The effect of the two connectivity motifs on the rest state vector is qualitatively different.
In the network with recurrent feedback connectivity (that is $\W{A}{}{G}{} = 0$), the rest state is:
\begin{equation}\label{eq:VBresFeedback}
\vV_{G}^\ast \,=\, \tau_B \, \tau_{G} \, \W{B}{}{G}{}.\pare{\cI_N \,+\, \eta\, \Gamma^{A}_{B} \,\Gamma^{B}_{A}}^{-1}.\vF,
\end{equation}
where $\cI_N$ is the $N$-dimensional identity matrix.
We see that the parameter $\eta$ acts on the rest state which is divisively modulated via the inverse matrix $\pare{\cI_N \,+\, \eta \, \Gamma^{A}_{B} \,\Gamma^{B}_{A}}^{-1}$. 
%
Note that, in the present context where $\Gamma^{A}_{B} =\Gamma^{B}_{A}$ and where these matrices are symmetric, the matrix 
$\cI_N \,+\, \eta\, \Gamma^{A}_{B} \,\Gamma^{B}_{A}$ is always invertible. 

For small weights, i.e. such that $\eta \|\Gamma^{A}_{B}\|_2 \|\Gamma^{B}_{A}\|_2 \ll 1$, the term $\cI_N$ dominates in the denominator and the rest state is controlled by $\W{B}{}{G}{}$, i.e. by the direct input from BCs to RGCs.

As $w^-$ increases, the influence of the feedback loop increases. With increasing $w^-$ the term $\eta\, \Gamma^{A}_{B} \,\Gamma^{B}_{A}$ grows, and when $\eta \|\Gamma^{A}_{B}\|_2 \|\Gamma^{B}_{A}\|_2 \gg 1$, the feedback influence dominates and the rest state decreases in amplitude as $w^-$ growths. The rest state of RGCs for different stimulus amplitudes decays like an hyperbola (Figure \ref{fig:1} C). Whatever the amplitude of $w^-$, the rest state remains positive, in contrast to the feedforward case discussed below. Even though the difference between rest states for different input amplitudes decreases with increasing  $w^{-}$, recurrent-feedback coupling still leads to a voltage response which scales with input amplitude and robustly maintains a response, even with strong inhibition.


The intermediate regime between "small" and "strong", $\eta \|\Gamma^{A}_{B}\|_2 \|\Gamma^{B}_{A}\|_2  \sim 1$ is further discussed in the supplementary material \ref{Sec:IntermediateEta}.
%
\\

For an RGC with feedforward inhibition the rest state vector is:
%
%
\begin{equation}\label{eq:VAres}
\vV_{G}^\ast \,=\, \tau_B \, \tau_{G} \, \pare{\W{B}{}{G}{} \,+\, \tau_A w^+ \W{A}{}{G}{}.\Co{B}{}{A}{}}.\vF.
\end{equation}
We see that the connections $\W{A}{}{G}{}$, multiplied by the time constant ($\tau_A$) of inhibition, act now subtractively on the rest state. The rest state now linearly scales with input amplitude (Figure \ref{fig:1} D) via  $\tau_B \, \tau_{G} \, \pare{\W{B}{}{G}{} \,+\, \tau_A w^+ \W{A}{}{G}{}.\Co{B}{}{A}{}}$.This expression acts as a scale "factor" here. As this "factor" is actually a matrix it can act differently depending on the how the stimulus projects to eigenvectors. Qualitatively, for small inhibitory weights $\W{A}{}{G}{}$, the scale factor is big, leading to a strong separation between rest states. As $\W{A}{}{G}{}$ increases, this separation becomes smaller because the scale factor  $\tau_B \, \tau_{G} \, \pare{\W{B}{}{G}{} \,+\, \tau_A w^+ \W{A}{}{G}{}.\Co{B}{}{A}{}}$ decreases. When the effect of inhibitory weights increases, the rest state will eventually become negative for those RGCs with index $k$ such that:
%
$$
\sum_{i=1}^N \pare{
\W{B}{i}{G}{k} \,+\, 
\tau_A w^+ \pare{\W{A}{i-1}{G}{k}+
\W{A}{i+1}{G}{k}}}.F_i < 0.
$$
Given that the RGC voltage will be rectified in this case, a modulation of the inhibitory strength below this threshold will not yield a change in the RGC spiking output.
\\


In response to a full-field impulse stimulus, the two motifs also qualitatively differ in their response shape (Figure \ref{fig:1} E). For this set of parameters, the feedforward inhibition leads to a biphasic response profile. The positive and negative phases come from the respective bipolar and amacrine inputs, with a delay corresponding to their respective characteristic times.
In contrast, reciprocal inhibition for the same set of parameters can produce, depending on the parameters, a multi-phasic profile into the impulse response. This multiphasic shape comes from damped oscillations in the system, due to the presence of complex frequencies \cite{Kartsaki2024}. The characteristic frequency of these oscillations depend on the parameter $\mu$, eq. \eqref{eq:mu} \cite{Souihel2021,Kartsaki2024}, which is governed by the connectivity weights of the system. Figure \ref{fig:1} \textbf{F,G} shows that the leading frequency increases with the strength of recurrent inhibition.




\subsection{Connectivity motifs give rise to different mechanisms for motion anticipation}

Anticipation is necessary to compensate for transmission delays in the signal transduction cascade \citep{Berry1999}.  In this section, we show how the inhibitory connectivity shifts the response peak to a moving object and can induce an anticipatory response to a moving bar.

\begin{figure}[h!]
\includegraphics[width=\textwidth,keepaspectratio] {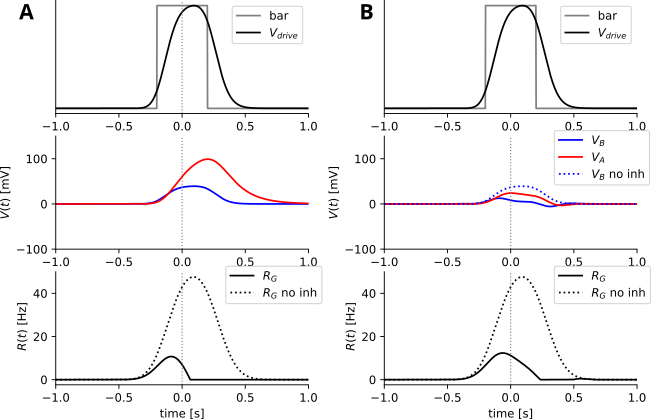}
\caption{\textbf{Response at each stage of the model, to a bar moving from left to right at $0.7$ $mm/s$, for the two connectivity motifs.} 
\textbf{A} \textit{Feedforward connectivity} evokes anticipation at the level of ganglion cells. 
\textbf{Upper panel:} Bar stimulus (grey) and  spatio-temporal convolution yields $V_{drive}$ (eq. \eqref{eq:V_drive}), which simulates the response of the OPL to a visual stimulus.
\textbf{Middle panel:}  BC (blue) and AC (red) voltage responses to the bar stimulus. 
\textbf{Lower panel:}  RGC firing rate in response to the bar stimulus when inhibition is present (solid black) compared to a purely excitatory feedforward network (dotted black). 
\textbf{B} \textit{Recurrent feedback connectivity} can evoke anticipation at the level of bipolar cells. Panels show same as in \textbf{A.}}
\label{fig:2}
\end{figure}

A moving bar with speed $v$, whose center is located at the left edge of the retinal network, $x = 0$ at time $t_0 =  0$, will be at the receptive field center $x_i$ of the downstream bipolar and ganglion cell $i$ at $t_{bar_i} = \frac{x_i}{v}$ (see Figure \ref{fig:s1} \textbf{A}).
 In our model, the moving bar stimulus first evokes a voltage response in the OPL, called $V_{drive}$ (eq. \eqref{eq:V_drive} and Figure \ref{fig:2} \textbf{A}). The peak in $V_{drive}$ lags behind the stimulus due to the convolution with the temporal kernel (see Figure \ref{fig:s1} \textbf{A,C}). Cellular integration at each stage of the downstream circuit would cause additional delays in the response.
In a purely excitatory network, the response peak of a BC and a RGC, at $t_{B_i}^{peak}$ and $t_{G_i}^{peak}$, respectively, will thus always lag behind the center of the bar with increasing delays at each stage of the network (see the dotted lines in Figure \ref{fig:2}).

Inhibition in the network can compensate for this delay by truncating the excitatory response. Inhibition strongly reduces the response amplitude of the network, and  at the same time shifts the response peak forward. If this peak shift becomes larger than the delay introduced by phototransduction and downstream integration, this corresponds to an anticipatory response. In the following two sections, we show how the mechanism behind the peak shift differs qualitatively and quantitatively through feedforward and recurrent feedback connectivity.  
%

\subsubsection{Feedforward inhibition implements motion anticipation at the level of Ganglion Cells}

With feedforward connectivity, phototransduction and integration delays persist in the second stage of the model, at the BCs level, and the response delay is increased further during integration into ACs membrane voltage (Figure \ref{fig:2},  \textbf{A}). RGCs then pool over both BCs and ACs, while AC inhibition arrives shortly after BC excitation. This delay leads to an initial excitation that is truncated when inhibition starts to rise, and soon fully suppressed. This yields an advancement of the response peak (Figure \ref{fig:2}  \textbf{A}, last row) which acts locally within the pooling radius of the RGCs. 
Thus, feedforward connectivity anticipates motion via a substractive truncation of excitatory inputs that acts locally in the network and yields a peak-advanced response at the level of RGCs. 

\subsubsection{Recurrent feedback implements motion anticipation at the level of Bipolar Cells}\label{Sec:MotionAnticipationBCs}
In the case of recurrent feedback connectivity, the first stage  of the model (OPL) remains unchanged. However, in the second stage, ACs now inhibit BCs  shortly after they begin to respond to the visual stimulus. A peak shift is thus induced already at this stage (Figure \ref{fig:2},  \textbf{B}).  
When the bar propagates, it stimulates BCs which transmit their response to connected cells via lateral connectivity. The presence of feedback loops between BCs and ACs creates oscillatory responses (as in Fig. \ref{fig:1}, F), which laterally travel as waves in \textit{both directions} (forward and backward), with a stronger effect forward (in the direction of the stimulus).  In this process, each BC can then be viewed as a wave source, triggered by the stimulus, emitting waves that interfere with other waves.  The response of a cell is thus a superimposition of its own response to the stimulus and to the lateral waves coming from other positions in the network. 
%
The influence of other cells exponentially decays with the distance, the shortest loops having therefore the strongest impact. Due to the difference in phases, the superimposition of waves creates an offset in the response peak, which can be in advance (anticipation) or delayed. In the settings chosen in this paper, the dominant effect of lateral connectivity is the suppression of the nearest neighbor cell, which truncates the response and is the primary cause of the anticipatory peak shift. 
The precise timing and amplitude of the response peak depends on the spatio-temporal Fourier spectrum, and also on the speed of the moving bar, which we will look at in the next section.



\subsection{Tuning to bar speed qualitatively differs in the two connectivity motifs}

The network response to a moving bar changes with speed. To illustrate this, we transform the temporal phase shift between response peak and bar position into a spatial measure. 
We calculate the peak shift for RGCs as $\delta X_{G_i}^{peak} = v \, \delta t_{G_i}^{peak} $ with  $\delta t_{G_i}^{peak} = t_{G_i}^{peak} - t_{bar_i}$.  

The lag induced by the temporal integration of $V_{drive}$ continuously increases with increasing speed (Figure \ref{fig:s1} \textbf{A,C}). This is because the bar spends less time in the receptive field with increasing speed (Figure \ref{fig:s1} \textbf{B}), but the "transduction delay" $\tau_{RF}$ stays constant. The gap between the two times thus increases with increasing speed. 

Similarly, the amplitude of the response to the moving bar decreases with increasing speed as the faster bar spends less and less time in the receptive field (Figure \ref{fig:s1} \textbf{D}). 

\subsubsection{Feedforward inhibition strongly anticipates slow speeds}

Feedforward inhibition strongly shifts the response peak for the slowest speed tested, so that $\delta X_{G_i}^{peak}$ is large. The peak shift $\delta X_{G}^{peak}$ then decreases with increasing bar speed. 
This can be explained as follows. Truncation always sets in when inhibition starts to increase, which in turn depends on when the bar enters the receptive field. The time the bar spends in the receptive field inversely scales with speed $v$ (Figure \ref{fig:s1} \textbf{B}), thus anticipation inversely scales to bar speed as well and is always maximal for the lowest speed (Figure \ref{fig:3} \textbf{C}). The peak shift via feedforward inhibition thus scales with the bar speed along an hyperbola. 

Slow bars spend more time in the receptive field and thus trigger strong responses. This results in a strong suppression of response amplitude for slow speeds, with amplitude increasing as the bar moves faster (Figure \ref{fig:3} \textbf{B}). For very fast speeds, inhibition is too slow to impact the excitatory response while the bar is in a cell's RF. From this point on, the network fails to anticipate, the amplitudes follow the scaling of $V_{drive}$ and start to decrease (Figure \ref{fig:s1} \textbf{D}).

\subsubsection{Recurrent inhibition tunes anticipation to bar speed}

\begin{figure}[h!]
\includegraphics[width=\textwidth,keepaspectratio] {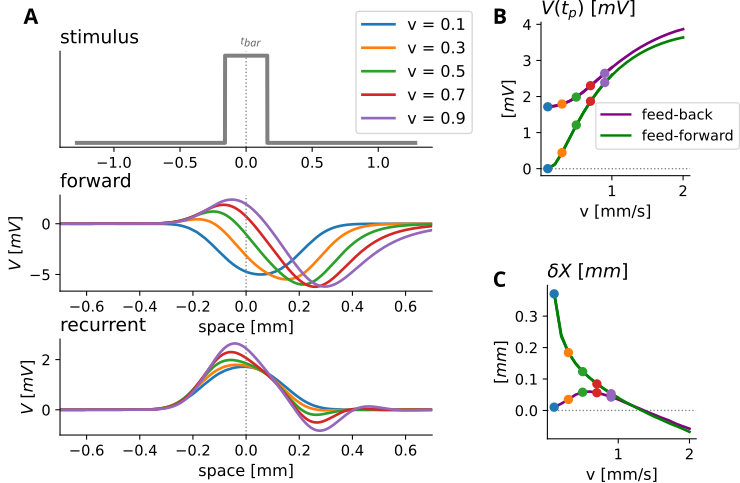}
\caption{\textbf{Feedforward and recurrent feedback inhibition result in qualitatively different tuning to bar speed.}
\textbf{A.} Upper: Bar stimulus. Below: Response traces of $V_{G}$ in the feedforward network (middle) and the recurrent feedback network (lower). Traces are plotted against the distance of spatial position of the bar center from the RF center at time $t$, motion from left to right. 
\textbf{B.} $\delta X^{i}_{G} $ plotted against bar speed.
\textbf{C.} Peak amplitude $V(t^{i}_{G})$ plotted against bar speed.}
\label{fig:3}
\end{figure}

In contrast, the response of the recurrent network is tuned to the bar speed, such that cells exhibit a preferred speed for which anticipation is maximal (Figure \ref{fig:3} \textbf{A, B}). 

This tuning arises because moving bars of different speeds induce different wave patterns in the network response, resulting in differently shaped responses (Figure \ref{fig:3} \textbf{A}, lower panel). Slow speeds trigger slow oscillating responses. As speed increases, responses oscillate faster. Faster oscillations lead  an earlier fist peak in the response, the peak shift is thus bigger. At the same time, the lag via the temporal integration delay increases with increasing speed.  Eventually  this lag becomes bigger than the increase in anticipatory shift, leading to a tuning curve with a maximally anticipated speed where the advancing shift is strongest relative to the lag. 

The peak amplitude tuning to different speeds in the recurrent feedback case is similar to the feedforward case, with exceptions for slow speeds (Figure \ref{fig:3} \textbf{B}). Here, the amplitudes remain at a higher level because the amplitudes of excitation and inhibition are recurrently coupled. 

Mathematically, the tuning of anticipation with respect to the bar speed can be understood via an analysis of the spatio-temporal Fourier transform of the network response. The speed to which anticipation is tuned is related to complex resonances in this Fourier transform. This will be presented in detail in another paper (in preparation). Here we present the main ideas. \\

Each point in the spatio-temporal Fourier spectrum has an amplitude, a wave number corresponding to a spatial periodicity, and a time frequency. 
In this model, the spatio-temporal Fourier transform of the response can be computed analytically and extended to complex wave numbers and frequencies.
In this context, the peaks appearing in the spectrum, for real space and time frequencies \cite{Kartsaki2024} correspond to resonances, i.e. poles in the complex domain \cite{cessac-sepulchre:04,cessac-sepulchre:06,cessac-sepulchre:07}. Thus, each pole corresponds to $2$ complex numbers. The first complex number (corresponding to a wave number)  reads $k^\ast_r + i k^\ast_i$, 
where $k^\ast_r$ fixes a scale for spatial oscillations, while $k^\ast_i$ fixes a characteristic space scale for exponential damping. Thus, this complex wave number determines a spatial response profile.  Likewise, the second complex number (corresponding to the time domain) is a complex frequency 
$\omega^\ast_r + i \omega^\ast_i$, 
where $\omega^\ast_r$ fixes temporal oscillations, while $\omega^\ast_i$ determines a characteristic time for exponential damping. Thus, this complex frequency determines a time response profile (see Figure \ref{fig:3} \textbf{A, B} and \cite{Kartsaki2024}).

It turns out that, in the response to a moving bar, the resonances/poles of the model depend on the speed $v$ as well as on connectivity and on the model parameters. For nearest neighbors connectivity the pole is unique (up to some symmetry).
It can then be shown that the $4$ quantities $k^\ast_r(v),k^\ast_i(v),\omega^\ast_r(v),\omega^\ast_i(v)$ determine the time to peak in the response and thereby the degree of anticipation.

\FloatBarrier 

\subsection{Tuning to bar speed depends on inhibitory strength}

The tuning between peak shift and bar speed via recurrent feedback coupling  depends on the parameter $\eta$ which characterizes the feedback intensity. A frequently observed property of circuits in the retina is that synaptic connections dynamically adapt during stimulation, such that connectivity weights are not static during presentation of a visual scene. We thus explored how the tuning to speed in the recurrent network is affected by connectivity weights. For this, we vary $w^-$ and thereby $\eta$. 

\begin{figure}[h!]
\includegraphics[width=\textwidth,keepaspectratio]
{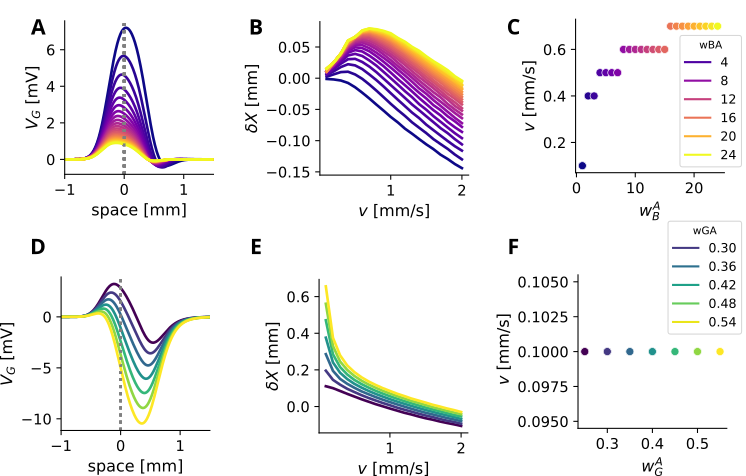}
\caption{ 
\textbf{Tuning to bar speed depends on inhibitory strength} with recurrent feedback (first row) and feedforward coupling (second row). The color legend applies to the whole row. \textbf{A}. Response for the recurrently coupled network to a moving bar at $v = 0.8 \, mm/s$ for a range of coupling strengths $w^-$ between 0 and 25 Hz. The bar is aligned at the receptive field center at 0.
\textbf{B}. Spatial peak shift plotted against bar speed for a range of coupling strengths  $w^-$ between 0 and 25 Hz. 
\textbf{C}. Speed with maximal peak shift plotted against coupling strengths. 
\textbf{D}. Response for the forward-inhibition network to a moving bar at $v = 0.8 \, mm/s$ for a range of inhibitory strengths $w^{A}_{G}$ between 0 and 0.5 Hz. The bar is aligned at the receptive field center at 0.
\textbf{E}.  Spatial peak shift plotted against bar speed for a range of forward strengths  $w^{A}_{G}$ between 0 and 0.5 Hz. 
\textbf{F}. Speed with maximal peak shift during feedforward inhibition. The fastest speed as always maximally anticipated.
}
\label{fig:4}
\end{figure}

The inhibitory strength of the recurrent feedback loop $w^-$ impacts the shape of the response: stronger coupling includes stronger and faster oscillation, which result in smaller response amplitude and in a stronger peak shift (\ref{fig:4} \textbf{B}). 
Over a range of speeds, stronger coupling also leads to a tuning towards faster preferred speeds (\ref{fig:4} \textbf{A}). 

In the feedforward connected network, stronger levels of inhibition lead to a stronger suppression of the response and thus to a smaller excitatory response. Stronger inhibitory weights lead to stronger overall anticipation with hyperbolic scaling. At low levels of inhibition, the scaling appears linear (\ref{fig:4} \textbf{C, D}). 


\FloatBarrier

\section{Discussion}
\label{Discussion}

\subsection*{Response properties to static and dynamic stimuli depend on connectivity patterns}

In a constantly moving visual environment, the retina has to reliably signal moving objects at various different speeds. At the same time, this organ is known to anticipate moving objects in order to compensate for transmission delays \citep{Berry1999,Johnston2015,Trenholm2013a, Trenholm2013b,Souihel2021}. In this study, we showed how two different inhibitory connectivity motifs both lead to an advancing shift in the peak response, corresponding to an anticipation mechanism. We highlight key differences between these connectivity motifs in the mechanistic implementation. When considering the rest state, feedback connectivity induces divisive inhibition while feedforward connectivity induces subtractive inhibition. When considering the anticipatory response to moving objects these two mechanisms induce strong differences in the peak shift and in the scaling of this shift with bar speed.


Feed-forward inhibition evokes motion anticipation at the level of ganglion cells via subtractive inhibition that vertically propagates through the network. Feedback connectivity, on the other hand, evokes motion anticipation at the level of bipolar and amacrine cells,  via an oscillatory response behavior, and triggers activity waves that laterally travel through the network 
The two motifs lead to different scalings between the shift of the response peak and the speed of the moving bar. While feedforward inhibition increases peak shifts for slow speeds, recurrent feedback coupling creates a preferred speed for which the peak shift is maximal.
The strength of inhibition thereby affects the relation between speed and anticipation. Increasing intensity of inhibitory recurrent feedback shifts this preference to faster speeds. 

Altogether, recurrent-feedback connectivity can exhibit effects which do not appear with a feed-forward architecture. It is more robust and maintains flexibility to  static inputs via divisive modulation. In addition, it permits to tune anticipation to object speed via adaptive oscillations.

 
\subsection*{Response properties to spatiotemporal stimuli can inform us about the upstream circuitry}

Here, we showed in a computational study how response properties to static and moving stimuli of RGCs differ due to the connectivity of the upstream network.
Examining response properties of different RGC types to these stimuli experimentally could thus hint us to how different circuits are wired.  
For example, a recent study shows that bipolar cell inputs anticipate motion \cite{dePiero2022} suggesting a mechanism that acts in the IPL already, such as the recurrent feedback motif presented here. However, one could also imagine a modified forward inhibition connectivity where amacrine cells laterally connect bipolars without reciprocal coupling, which would also lead to anticipation at the bipolar level, but with different speed scaling properties.
Anticipation scaling with speed has been shown behave qualitatively different in different RGC types \cite{Berry1999}. Some scale more hyperbolic, as would result from feed-forward inhibition in our model, while others show a maximally anticipated speed, as in our recurrent-feedback simulations. Yet others exhibit constant anticipation across a range of speed, which could indicate more elaborated mechanisms (see below). 
It thus seems likely that different cell-types could employ different mechanisms for motion processing. A combination of feedforward and recurrent feedback inhibition could also give rise to more elaborate tuning curves which we did not study here.

Moving objects of different speeds also evoke different tuning in response amplitude in different cell-types \cite{Matsumoto2025, Summers2022}. However, the two connectivity motifs studied here do not provide testable distinction between the two connectivity motifs in terms of amplitude scaling.

With feedforward connectivity, response amplitudes to low speeds can get very small and disappear with sufficiently large inhibition. This could partially explain why some cells seem to not respond at all to low frequencies. Feedback inhibition, on the other hand, prevents responses from becoming too small at slow speeds and could maintain a broader amplitude tuning.

Amacrine cells have been shown to play central roles in many retinal computations \cite{Gollisch2010, Jarsky2011, Diamond2017, Franke2017a, Franke2017b, Korympidou2024}, but the precise role of their diversity and complex circuitry are only starting to be understood \cite{Huang2019, Calbiague2025}.
With a detailed understanding of how different connectivity motifs shape retinal response properties, analyzing these properties in retinal cell-types might help to inform us about their upstream circuitry.  For example, oscillatory response behavior and spatiotemporal frequency preferences could extend a receptive field characterization. 


\subsection*{Speed representations via dynamic feedback circuits}
Recent studies suggest that dynamic adaptation of retinal circuits play important roles in shaping retinal computations \cite{Kastner2011, Yu2022, Huang2022, Deng2024}, and that dynamic inhibitory synapses can lead to a better representation of temporal stimulus properties in the retina \cite{Ebert2024}. 
The retinal network could also adapt to a moving bar for a representation of bar speed by dynamically tuning its connectivity weights via short term plasticity. 
Such an adaptation to a more complex input feature could allow a stable anticipatory signal across a range of speed, as observed in \cite{Berry1999}. One could imagine that inhibitory synapses in the network depress in response to the strong activation by slow speed, which in turn would shift maximal anticipation towards slow speed, thereby stabilizing anticipation across a range of speeds.
In addition, especially the recurrent-feedback circuit is not only characterized by the connectivity weight but by the product $w^- w^+ \tau_A \tau_B$. Therefore, modifying the characteristic times can also impact the response to moving objects and could also occur via dynamically adapting synapses.

The fact that the retina can anticipate its future inputs supports the idea that its pursues a predictive coding strategy to efficiently encode an entire visual scene \cite{Barlow1961, Srinivasan1982, Rao1999, Hosoya2005, Salisbury2016}. The predictive coding theory posits that neuronal circuits not only form predictions about future inputs, but also emit error signals when these are not matched \cite{Sederberg2018, Hoshal2024, Barlow2001}. In line with this idea, it has been observed that retinal ganglion cells signal unexpected motion in a visual scene, such as onset \cite{Chen2013}, reversal \cite{Schwartz2007c, Chen2014} and interruption \cite{Ding2021}. How these different aspects are implemented in neuronal circuits is however not well understood. In the future, it will be interesting to study how different mechanistic implementations for distinct predictive coding strategies could interact to enable the retina with its rapid and efficient information processing.

\subsection*{Limitations}

In this study, we simulated all cells as point neurons, which are divided into bipolar, amacrine and ganglion cells by the network architecture and time constants but otherwise share response properties. We thus cannot reference the circuits analyzed here to specific and experimentally observed retinal circuits but rather focus on general effects of feedback versus feedforward connectivity. 
For example, it is known that A17 amacrine cells form reciprocal synapses with rod-bipolar cells that affect local temporal response properties \cite{Grimes2010}, but whether these recurrent loops also act on a more global level is not clear.

It is not clear as well to which extent the inhibitory cells in our network could also resemble horizontal cells. The more complicated feedback mechanisms between horizontal cells and photo-receptors are less likely to induce  oscillations, as these mechanisms do not rely on synaptic transmission \cite{Kramer2015}. 
Feedforward inhibition however might already play a role in anticipation at the first synapses in the visual system. 
Inhibitory neurons in our model could thus partially represent horizontal cells, mainly via feed-forward inhibition.

One important difference between recurrent feedback and feedforward connectivity is that recurrence introduced a divisive modulation of the voltage response, which can have important consequences for neural computations \citep{Carandini2012}. While we showed how this divisive modulation acts on the rest state in response to static inputs, its effect in response to dynamic stimuli - where transient effects dominate - remains to be studied in detail.


\section{Methods}\label{Methods}

\subsection{Parameter Calibration}

We set receptive field sizes, number of cells and cell spacing and ganglion cell pooling width to the values used in a previous model for with spatiotemporal convolution for motion detection \cite{Souihel2021,Kartsaki2024}, studies themselves based upon \cite{Berry1999,Chen2013}. We then manually optimized time constants and connectivity weights, and scale factors such that the model responses qualitatively match experimentally recoded responses, provided by Thomas Buffet and Oliver Marre from the Vision Institute in Paris (Figure \ref{fig:param_opt}). 

The firing rate in response to a moving bar with width $b = 160$ $\mu m$ and $v = 0.7$ $mm/s$ was obtained via multi-electrode array recodings. The spatial and temporal receptive field of the cell was determined via spike-triggered-average (STA) computation with  white noise stimulation. The response to the moving bar of the cell was aligned to 0 at the time where the moving bar was located over the experimentally defined spatial receptive field center. We then compared the voltage response of a simulated RGC, in response to a full-field impulse stimulus, to the  experimentally obtained temporal STA of the cell. Then, we compared the predicted firing rate of the model, in response to a moving bar of the same speed, to the recorded experimental response in order to optimize connectivity weights and time constants.


\clearpage
\subsection{Parameter Values}

\begin{table}[h!]
\centering
 \begin{tabular}{||c c c||} 
 \hline
 Parameter & Value & Unit  \\ [0.5ex] 
 \hline\hline
 
 $\sigma_{B}$             & 0.05       & $mm$          \\ 
 $\sigma_{G}$             & 0.065      & $mm$          \\ 
 $\delta$                 & 0.005       & $mm$         \\ 
 $\tau_{RF}$              & 0.04        & $s$          \\ 
 $\tau_{B}$               & 0.08        & $s$          \\
 $\tau_{A}$               & 0.15        & $s$          \\
 $\tau_{G}$               & 0.01        & $s$          \\
 
 $w^{-}$                  & -10.0       & $Hz$         \\ 
 $w^{+}$                  & 10.0        & $Hz$         \\ 
 $w^B_G$                  & 0.8         & $Hz$         \\ 
 $w^A_G$                  & -.4         & $Hz$         \\ 
 $s_{G}$                  & 5           & $Hz mV^{-1}$ \\ 
 $\theta_{G}$             & 0.0         & $mV$         \\ 
 $a_{mV}$                 & 20.0        & $nS^{-1}$    \\ 
 $dt$                     & 0.001       & $s$          \\
 $stimulus intensity$     & 1.0/0.1     & $pA$         \\ 

 \hline
 \end{tabular}
 \caption{Parameter values used in simulations, if not stated otherwise in the text.}
 
 \label{tab:parameterc4}
\end{table}

\section*{Acknowledgements}

We thank Olivier Marre, Philipp Berens and Adrian Palacios for helpful discussions about the model. We thank Thomas Buffet for  letting us to use his data to find realistic parameters for our model. This work has been funded by a PhD fellowship from the interdisciplinary Institute for Modeling in Neuroscience and Cognition (NeuroMod) of Université Côte d’Azur to S.E., funded by the National Research Agency (ANR-15- IDEX-01) and, by an ERC grant (No 101045253, DEEPRETINA) to O.M. and ERC grant (No 101039115, NextMechMod) to P. B. . 

\section*{Author Contributions}
S.E. and B.C. designed the study. S.E. implemented the model and did numerical simulations, B.C. did analytical computations.  S.E. and B.C. wrote the paper.

\section*{Code Availability}
The code used for the computational model is accessible on \href{https://github.com/simoneeb/motion_anticipation_network}{github} and will be set tot public after publication.

\section*{Competing Interests}
The Authors declare no competing interests. 

\bibliographystyle{abbrv}
\bibliography{bibliography.bib}
\newpage
\newcommand{\beginsupplement}{%
        \setcounter{table}{0}
        \renewcommand{\thetable}{S\arabic{table}}%
        \setcounter{figure}{0}
        \renewcommand{\thefigure}{S\arabic{figure}}%
     }
     
\captionsetup[figure]{font=small,skip=0pt}
\section*{Supplementary Material}
\beginsupplement

\subsection*{Model Equations} \label{Sec:Equations}

In order to simulate the integration of lateral inputs, we implement the retinal network as a dynamical system, rather than as a cascade of convolutions \cite{Maheswaranathan2018}. The model can be described as follows.

We consider a 1-D moving bar stimulus $s(t)$  (in $pA$) with a speed $v$ (in $mm/s$) and width $2b$ (in $mm$). The stimulus is described as follows:

 \begin{equation}
  s(x,t) =
    \begin{cases}
      1,  &\text{if $-b +vt \leq x \leq b+vt $ ; } \\
      0,  &\text{otherwise.}
    \end{cases}
\label{eq:s(t)}
\end{equation}

We simulate the voltage input from the OPL to a bipolar cell $i$ with a receptive field center located at $x_{i}$ via a spatiotemporal convolution:
\begin{equation}
    V_{i}^{drive} (t) = a_{mV} \int_{-\infty}^{t}\int_{0}^{D} \cK(x_{i}-x,t-u) s(x,u) dxdu
\label{eq:V_drive}
\end{equation}
where the kernel is separated:
\begin{equation}
    \cK(x,t) = \cK_{T}(t)\, \cK_{S}(x),
\label{eq:K}
\end{equation}
i.e. it consists of the product of temporal ($\cK_{T}$) and a spatial ($\cK_{S}$) profile. The temporal profile is given by:
\begin{equation}
    \cK_{T}(t) = \frac{t}{\tau_{RF}^{2}} e^{-\frac{t}{\tau_{RF}}} 
\label{eq:kt}
\end{equation}
where $\tau_{RF}$ is the characteristic integration time of the OPL. The spatial kernel is a Gaussian: 
\begin{equation}
   \cK_{S}(x) = a_{mV} \, e^{-\frac{(x-x_{i})^{2}}{2\sigma^{2}_{B}}} 
\label{eq:kx}
\end{equation}
where $\sigma_{B}$ parametrizes the size of the receptive field center (in $mm$) of cell $i$ with position $x_{i}$ and $a_{mV}$ is a scale factor  with unit $nS^{-1}$ to transfrom the OPL output into $mV$.

We then consider a retinal network spanning a 1-D plane with N = 512 bipolar cells positioned at $x_{i}$ and N = 512 amacrine cells with the same spatial location. Cells cover a distance $D = 2.56$ $mm$ and are spaced by $\delta = 0.005$ $mm$. Each cell is characterized by its membrane potential $V_{B_i}$ and $V_{A_j}$ respectively. The dynamics of the cells is ruled by the dynamical system: 

 \begin{equation}
    \begin{cases}
     &\frac{dV_{B_i}}{dt} = -\frac{V_{B_i}}{\tau_{B}} - w^{-} \, \sum_{j=1}^{N} \Gamma^{A_j}_{B_i} V_{A_j} +F_{i}(t), \\
%
%
     &\frac{dV_{A_j}}{dt} = -\frac{V_{A_j}}{\tau_{A}} + w^{+} \,\sum_{i=1}^{N} \Gamma^{B_i}_{A_j}V_{B_i},
    \end{cases}
\label{eq:dynamical_systemc4}
\end{equation}

The connectivity is simulated via the matrices $\Gamma^{B}_{A}$ and $\Gamma^{A}_{B}$, which define the connections from BCs to ACs and from ACs to BCs, respectively.  Each BC $i$ projects onto ACs $j = i-1$ and $j = i+1$ and vice versa such that $\Gamma^{B_{i}}_{A_{i-1}} = \Gamma^{B_{i}}_{A_{i+1}} = 1$. All other entries are set to $0$. We assume null boundary conditions. Connections from BCs to ACs are excitatory and have a synaptic weight $w^{+} \geq 0$ while connections from ACs to BCs are inhibitory and have a synaptic weight $-w^{-} \leq 0$. The connectivity between BCs and ACs is symmetric, $\Gamma^{B}_{A} = \Gamma^{A}_{B}$.  This is mathematical choice commented in \citep{Kartsaki2022} allowing to avoid linear instabilities in the system, as the eigenvalues of the linear system always have a negative real part with this condition. 

The term:
\begin{equation}
 F_{i}(t) = \frac{V_{i}^{drive}}{\tau_{B}} + \frac{dV_{i}^{drive}}{dt},
\label{eq:F}
\end{equation}
is the stimulus driven input into BCs. It takes this form to ensure that $V_{Bi} = V_{i}^{drive}$ in the absence of ACs coupling


Finally, a layer of $N$ = $512$ Retinal Ganglion cells (RGCs) is added, obeying the differential equation: 
\begin{equation}\label{eq:Vg}
 \frac{dV_{G_k}}{dt} 
 = -\frac{V_{Gk}}{\tau_{G}} + \sum_{i=1}^{N} \W{B}{i}{G}{k} V_{B_i}(t)
 + \sum_{j=1}^{N} \W{A}{j}{G}{k} V_{A_j}(t),
\end{equation}
%
where each RGC $k$ pools over the bipolar cell layer with Gaussian weights $\W{B}{i}{G}{k}$, centered at the RGC's position $x_{k}$ (same as BC and AC position), and a width  $\sigma_{G}$ in $mm$. The scale factor $w^{B}_{G}>0$ (in Hz) determines the overall strength of synapses from bipolar to ganglion cells.
\begin{equation}\label{eq:wxb}
\W{B}{i}{G}{k}\,=\, w^{B}_{G} \, e^{-\frac{(x_i-x_k)^{2}}{2{\sigma_{G}^{2}}}}.
\end{equation}

In the feedforward network, each RGC pools as well over ACs with the same distribution. The synaptic strength is scaled by $w^{A}_{G}<0$ (in Hz):

\begin{equation}
\W{A}{j}{G}{k}\,=\, w^{A}_{G} \, e^{-\frac{(x_j-x_k)^{2}}{2{\sigma_{G}^{2}}}}.
\label{eq:wxa}
\end{equation}

In the last step, the RGC voltage $V_{G_k}$ is transformed into a firing rate $R_{Gk}$:
\begin{equation}
    R_{Gk} = N(V_{Gk},\theta_{G}).
\label{eq:Rg}
\end{equation}
via the piecewise-linear function: 
 \begin{equation}
  N(V) =
    \begin{cases}
      s_{G} (V-\theta), &\text{if $ V \geq \theta$ ;}\\
      0,  &\text{otherwise.}
    \end{cases}
\label{eq:N}
\end{equation}

For simulations with feedforward connectivity, we set $w^{-} = 0 $ to remove feedack inhibiton. For simulations for feedback connectivity, we set  $w^{A}_{G} = 0$.

\subsection*{The intermediate regime between "small" and "strong" $\eta$} \label{Sec:IntermediateEta}

Here, we come back to the equation \eqref{eq:VBresFeedback} for the rest state in the case of feedback inhibition. In the main text, we have discussed the extreme cases where $\eta$, the feedback loop intensity, is either very small ($\eta \|\Gamma^{A}_{B}\|_2 \|\Gamma^{B}_{A}\|_2  \ll 1$) or very large ($\eta \|\Gamma^{A}_{B}\|_2 \|\Gamma^{B}_{A}\|_2  \gg 1$). Here, we briefly discuss the intermediate case where 
$\eta \|\Gamma^{A}_{B}\|_2 \|\Gamma^{B}_{A}\|_2  \sim 1$. If we decompose the input over the eigenbasis $\vpsi_n$ of $\Gamma^{A}_{B} \,\Gamma^{B}_{A}$, i.e. $\vF=\sum_{n=1}^N F_n \vpsi_n$ then the rest state reads, in this basis,
$\vV_G=\tau_G \sum_{n=1}^N \frac{F_n}{1+\eta \kappa_n^2} \W{B}{}{G}{} \vpsi_n$.
The terms $\frac{F_n}{1+\eta \kappa_n^2}$
have a very different amplitude in this regime, due to the denominators $\frac{1}{1+\eta \kappa_n^2}$.
The largest denominator corresponds to the eigenvalue index $n$ such that $1+\eta \kappa_n^2$ is the smallest.
Thus, even if the input is spatially homogeneous, the rest state will show a spatial inhomogeneity favoring the space scale 
$\frac{N\delta}{n}$
corresponding to the eigenvector $\vpsi_n$.

\newpage

\section*{Supplementary Figures}

\begin{figure}[h!]
\includegraphics[width=\textwidth,keepaspectratio]
{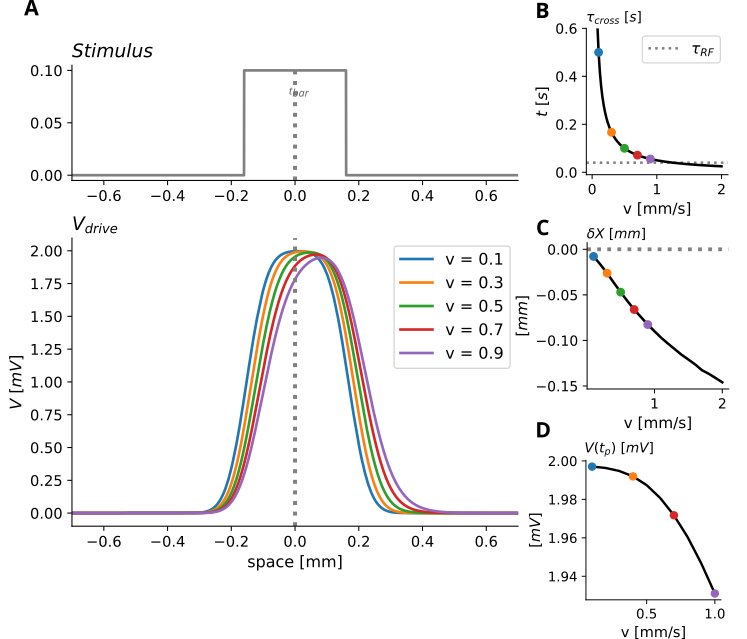}
\caption{\textbf{Photo-transduction layer introduces a lag between response peak and bar middle which increases with the speed of the bar} 
\textbf{A.} Upper: Bar stimulus. Lower: Response traces  of $V_{drive}$ to bars moving at speeds between 0.1 and 1.0 mm/s. Traces are plotted against the distance of spatial position of the bar center from the RF center at time $t$, motion from left to right. 
\textbf{B.} $\tau_{cross}$ plotted against bar speed. $\tau_{RF}$ is indicated by the grey dotted line.
\textbf{C.} $\delta X^{i}_{drive} $ plotted against bar speed.
\textbf{D.} Peak amplitude $V(t^{i}_{drive})$ plotted against the bar speed.}
\label{fig:s1}
\end{figure}
\FloatBarrier

\begin{figure}[h!]
\includegraphics[width=\textwidth,keepaspectratio]
{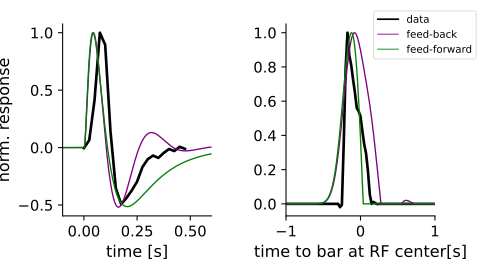}
\caption{\textbf{Model responses optimized to experimental data from of a recorded RGC which anticipates motion. }
\textbf{A.} Experimental temporal STA of the RGC and impulse response fit of the model (Courtesy of Olivier Marre and Thomas Buffet).
\textbf{B.} Firing rate in response to moving bar at 0.7 mm/s and simulations of the network, aligned at t=0 to when the bar is at the center of the receptive field (dotted line). Amplitudes are normalized for comparison.}
\label{fig:param_opt}
\end{figure}
\FloatBarrier

\end{document}